# Better Recommendations: Validating AI-generated Subject Terms Through LOC Linked Data Service

Kwok-leong Tang[a]* and Yi Jiang[b]

[a]*Department of East Asian Languages and Civilizations, Harvard University, Cambridge, USA;*
[b]*Jerome Hall Law Library, Indiana University, Bloomington, USA*

Email of corresponding author: kwokleongtang@fas.harvard.edu

# Better Recommendations: Validating AI-generated Subject Terms Through LOC Linked Data Service

Abstract: This article explores the integration of AI-generated subject terms into library cataloging, focusing on validation through the Library of Congress (LOC) Linked Data Service. It examines the challenges of traditional subject cataloging under the Library of Congress Subject Headings (LCSH) system, including inefficiencies and cataloging backlogs. While generative AI shows promise in expediting cataloging workflows, studies reveal significant limitations in the accuracy of AI-assigned subject headings. The article proposes a hybrid approach combining AI technology with human validation through LOC's Linked Data Service, aiming to enhance the precision, efficiency, and overall quality of metadata creation in library cataloging practices.

## Introduction: Validating AI-Generated Subject Terms

This article proposes an innovative approach to validate AI-generated subject terms through the Library of Congress (LOC) Linked Data Service. By leveraging this validation framework, libraries can significantly accelerate their cataloging processes while maintaining metadata quality standards. We begin by exploring the current landscape of generative AI adoption in subject heading suggestions and analyzing persistent challenges in accuracy and consistency. The article then examines various methodological approaches to improve the precision of AI-generated subject terms, with particular attention to controlled vocabulary alignment and semantic validation techniques. Finally, we present a practical solution for enhancing the accuracy of AI-suggested subject headings through LOC Linked Data Service integration, demonstrating its implementation in three production-ready cataloging tools that streamline workflows while preserving librarian expertise in metadata creation.

## Challenges in Subject Cataloging Practices

Subject cataloging remains a crucial component across all cataloging work. The United States Library of Congress operates the extensive controlled vocabulary thesaurus known as

Library of Congress Subject Headings (LCSH), which serves as the main subject cataloging system. The standardized and controlled vocabulary provided by this system serves as a foundational element of bibliographic control for organizing and retrieving materials in libraries. As one of the most widely used controlled vocabularies, LCSH enhances information access in several key ways. LCSH brings consistency to library collections by categorizing topics into logical arrangements and by controlling synonyms, variant spellings, and homographs.[1] The system ensures standardization throughout libraries and therefore enables users to employ consistent search methods. LCSH improves subject retrieval by uncovering hidden aspects of materials that titles and tables of contents don't reveal, while delivering precision that proves vital for research. Its controlled vocabulary also enables cross-lingual searching, which supports access to non-text and multilingual resources.[2] Furthermore, LCSH provides bibliographic organization of materials, which results in more efficient information retrieval processes and better supports scholarly research.[3]

      For library catalogers, assigning subject headings is actually a systematic work, the purpose of which is to make the subject marking of books or materials both accurate and consistent. They will first carefully analyze what the book or material is about and determine its focus and main topics. Then, they will look up the established headings in the subject heading authority files and try to choose the most appropriate and specific words or phrases to describe the content. According to the basic rules of LCSH, at least one subject heading that summarizes the main content should be assigned when cataloging. The "20% rule" requires that the selected subject heading should represent at least one-fifth of the content of the book. If the content is complex and one heading cannot explain it clearly, multiple headings can be used to represent it. Generally speaking, at most three or four subject headings will be selected, which is complete but not too lengthy. In addition, detailed subdivisions (such as geographic, chronological, form, etc.) can be added to make the subject headings more specific. However, the use of all subject headings must comply with the LCSH policy guide as detailed in the "Subject Heading Manual"

---

[1] "Process for Adding and Revising Library of Congress Subject Headings," accessed June 17, 2025, https://www.loc.gov/aba/cataloging/subject/lcsh-process.html.

[2] "On Outdated and Harmful Language in Library of Congress Subject Headings – News from Columbia's Rare Book & Manuscript Library," accessed June 17, 2025, https://blogs.library.columbia.edu/rbml/2021/10/19/on-outdated-and-harmful-language-in-library-of-congress-subject-headings/.

[3] Vanda Broughton, *Essential Classification*, Second edition. (Chicago: ALA Neal-Schuman, an imprint of the American Library Association, 2015), 111–12.

(SHM). If catalogers can't find a suitable subject heading, they can also propose new headings according to SHM principles. Through this process, the use of subject headings not only allows library resources to be better organized but also makes it easier for users to find the information they need.

However, due to the complexity of the LCSH system itself, subject cataloging often becomes a persistent challenge for professional catalogers in actual work. For example, adding subdivisions while using the LCSH seems to be just a trivial inclusion of supplementary information, but in fact, it is a time-consuming process that requires careful checking of instructions and accuracy. Another example is the use of "Free-Floating Subdivisions" (FFS) and Pattern Subdivisions. Although these types of subdivisions are more flexible in design, they must still be used in strict accordance with the scope of application listed in the Subject Headings Manual (SHM).  As a result, catalogers often have to check each combination of subject terms and subdivision terms one by one and manually update records to ensure consistency. Sometimes new subdivision terms are proposed, which must be reviewed before they can be approved, further lengthening the process. Although the specific time spent varies from person to person, most catalogers feel that dealing with subject cataloging alone takes up a large part of their work time, especially when they are also responsible for NACO-related (Name Authority Cooperative Program) work. Such operations that require constant checking of rules are not only time-consuming but also prone to errors.

For the reasons discussed above, subject cataloging based on the LCSH system has become one of the main reasons why many libraries and archives suffer from significant cataloging backlogs. The problems caused by such backlogs cannot be ignored. It not only affects the operation of an institution, but also directly affects the user experience. The most direct consequence is the delay in the listing of new collection items. Many newly purchased materials will not appear in the catalog for a certain period of time because they have not been cataloged in time, and users will not be able to search for or borrow them. This kind of "invisible collection" problem means that many valuable resources are actually "sealed" and cannot be accessed for teaching, research, or general knowledge acquisition. At the same time, these uncataloged materials also take up physical space without contributing to the library's mission, causing inefficient use of storage and shelf space. From a financial perspective, the backlog also reduces the return on institutional investment: a library spends a large amount of its budget to purchase new materials, but it cannot make them discoverable and accessible to users.

The emergence of generative AI has brought new hope to address these challenges. The integration of AI technology into library cataloging has gained traction with the advancement of large language models (LLMs). Custom GPTs have emerged as tools capable of generating MARC records based on provided content. After some data training and fine-tuning with these models, they have demonstrated high accuracy in extracting key bibliographic details such as title, author, and publisher information, even with foreign language titles that need Romanization.[4] However, it has also been noted that while these AI-generated records perform well in basic metadata fields, the assignment of subject headings requires significant human oversight. This observation aligns with the study led by Brzustonwicz, which found that although ChatGPT-generated records were comparable in accuracy to manually created records, differences emerged in the assignment of subject access points.[5] These discrepancies highlight the continued necessity of automatic validations and human reviews in ensuring precise subject representation and nuanced classification within LCSH. In short, there are now powerful AI-based tools that demonstrate potential as a solution to expedite the subject cataloging while at the same time suffer from their respective limitations and drawbacks. In the present work, we analyze existing explorations of AI-assisted subject cataloging and further propose an innovative and highly automated approach to validate AI-generated subject terms through the Library of Congress (LOC) Linked Data Service.

**Accuracy of AI-Generated Subject Terms**

Recent studies show mixed results regarding the accuracy of AI-generated subject headings. Large Language Models (LLMs) like ChatGPT can produce Library of Congress Subject Headings (LCSH) that sometimes align with those chosen by professional catalogers, but often with significant gaps. For example, a 2024 experiment used ChatGPT to assign LCSH terms to 30 electronic theses and dissertations and found that only about half of the AI-generated headings were both valid LCSH terms and sufficiently specific. The model frequently struggled with complex multi-part headings and subdivisions, outputting some terms that were not authorized LCSH entries (e.g. single-word or colloquial topics).[6] Another evaluation in 2025 tested three AI chatbots (ChatGPT, Google's Gemini, and Microsoft Copilot) on assigning classification codes and LCSH. It reported overall poor performance, especially for classification numbers, and "though subject heading assignment was also poor, ChatGPT showed more promise here". Common errors included choosing overly broad subjects or entirely wrong topics.[7] These findings echo earlier work with other AI tools. A 2021 prototype called "Kratt" indexed books with terms from the Estonian Subject Thesaurus; while it could generate some relevant keywords, librarians rated the quality as unsatisfactory, citing many inaccurate or missing terms. Even so, casual library users in that study found the AI-chosen terms somewhat useful for discovery, suggesting that what experts deem inaccurate might still help end-users to an extent.[8] Overall, the literature indicates that current generative AI systems alone do not reach the high accuracy standards required for subject cataloging, with reported precision or F1 scores far below professional cataloging levels (e.g. 26–35% alignment with human-assigned LCSH in one evaluation).[9] These accuracy limitations underscore the need for human validation – as one study concludes, "AI chatbots do not show promise in reducing time and effort associated with subject cataloging at this time, [underscoring] the continuing importance of human expertise and oversight."[10] However, researchers are optimistic that accuracy will improve as models are refined, larger training corpora are used, and hybrid human–AI workflows are developed.[11]

---

[4] Yi Jiang et al., "Exploring the Potential Applications of Generative AI in East Asian Librarianship: Use Cases, Reflections, and Inspirations," *Journal of East Asian Libraries* 2024, no. 179 (October 29, 2024), https://scholarsarchive.byu.edu/jeal/vol2024/iss179/5.

[5] Richard Brzustowicz, "From ChatGPT to CatGPT: The Implications of Artificial Intelligence on Library Cataloging," *Information Technology and Libraries* 42, no. 3 (September 18, 2023), https://doi.org/10.5860/ital.v42i3.16295.

**Efficiency Gains in Cataloging Workflows**

---

[6] Eric H. C. Chow, T. J. Kao, and Xiaoli Li, "An Experiment with the Use of ChatGPT for LCSH Subject Assignment on Electronic Theses and Dissertations," *Cataloging & Classification Quarterly* 62, no. 5 (July 3, 2024): 574–88, https://doi.org/10.1080/01639374.2024.2394516.

[7] Brian Dobreski and Christopher Hastings, "AI Chatbots and Subject Cataloging: A Performance Test," *Library Resources & Technical Services*, January 1, 2025, https://trace.tennessee.edu/utk_infosciepubs/485.

[8] Marit Asula et al., "Kratt: Developing an Automatic Subject Indexing Tool for the National Library of Estonia," *Cataloging & Classification Quarterly* 59, no. 8 (October 25, 2021): 775–93, https://doi.org/10.1080/01639374.2021.1998283.

[9] Isabel Brador, "Could Artificial Intelligence Help Catalog Thousands of Digital Library Books? An Interview with Abigail Potter and Caroline Saccucci | The Signal," webpage, The Library of Congress, November 19, 2024, https://blogs.loc.gov/thesignal/2024/11/could-artificial-intelligence-help-catalog-thousands-of-digital-library-books-an-interview-with-abigail-potter-and-caroline-saccucci.

[10] Dobreski and Hastings, "AI Chatbots and Subject Cataloging."

[11] Asula et al., "Kratt"; Chow, Kao, and Li, "An Experiment with the Use of ChatGPT for LCSH Subject Assignment on Electronic Theses and Dissertations."

Despite accuracy concerns, AI-driven subject term validation offers notable efficiency benefits. Speed and throughput improvements are a major motivation for integrating AI into cataloging. In the National Library of Estonia's "Kratt" project, the AI could assign subject keywords to a book in about 1 minute – making it 10–15 times faster than a human cataloger on average.[12] Even if the AI's suggestions weren't perfect, this drastic reduction in processing time highlights the potential for handling large backlogs of materials. Similarly, experiments at the Library of Congress (LC) targeted a growing backlog of uncatalogued e-books by seeing if machine learning could generate bibliographic metadata at scale. The AI models in LC's trial produced subject heading suggestions that catalogers could quickly accept or refine through a human-in-the-loop interface.[13] This approach is promising because it shifts librarians' time from creating headings to reviewing AI-suggested headings, effectively speeding up the workflow. In the experiment led by Chow, Kao, and Li, using ChatGPT to draft LCSH for theses, the cost was only about $0.25 and 3 minutes of AI processing for 30 documents – far faster and cheaper than manual cataloging. The authors note that refining an existing (even if imperfect) AI suggestion is "less daunting than constructing new subject headings from scratch".[14] In practice, this means junior catalogers or those pressed for time can use the AI's output as a first draft, then concentrate their expertise on correcting and enhancing those suggestions. Early field reports also describe productivity gains: a 2023 public library experiment found that GPT-4 greatly accelerated tasks like aligning taxonomy terms to user search behavior, turning a time-consuming analysis into an automated process.[15] Across these studies, AI is seen as a labor-saving assistant tackling the bulk of routine subject analysis instantly, so that librarians can handle higher-level decision making. It's important to note that the net efficiency gains depend on workflow design. When AI suggestions are integrated via easy-to-use tools (e.g. auto-suggestion interfaces or batch processing), catalogers can validate and finalize records much faster than doing everything manually.[16] If, however, the AI output requires extensive correction, the time savings diminish. The consensus in recent literature is that semi-automated cataloging,

---

[12] Asula et al., "Kratt."
[13] Brador, "Could Artificial Intelligence Help Catalog Thousands of Digital Library Books?"
[14] Chow, Kao, and Li, "An Experiment with the Use of ChatGPT for LCSH Subject Assignment on Electronic Theses and Dissertations."
[15] M. Ryan Hess and Chris Markman, "Beyond the Hype Cycle: Experiments with ChatGPT's Advanced Data Analysis at the Palo Alto City Library," *The Code4Lib Journal*, no. 58 (December 4, 2023), https://journal.code4lib.org/articles/17867.
[16] Brador, "Could Artificial Intelligence Help Catalog Thousands of Digital Library Books?"

with AI generating candidate terms and humans validating them, can significantly boost cataloging throughput – potentially cataloging collections that would otherwise remain unprocessed.[17] Libraries can thus leverage AI for basic term generation to improve efficiency, provided that robust validation steps are in place to maintain quality.[18]

**Enhancing Subject Term Recommendations with AI**

AI-driven tools are also being used to enhance subject term recommendations, complementing human catalogers' judgment and potentially improving metadata quality and consistency. One benefit is that LLMs, trained on vast text corpora, can surface relevant topics and terminology that a cataloger might not immediately think of. For instance, ChatGPT demonstrated an ability to draw on its internalized knowledge of LCSH and MARC records, sometimes suggesting legitimate LCSH headings that were closely similar to professional choices. In the experiment on thesis cataloging, the AI typically generated 3–5 subject headings per document (often including standard subdivisions). Catalogers noted that many of these were valid and on-topic, providing a useful starting list. AI can also recommend additional terms that, while not authorized LCSH, reflect important concepts in the material. Rather than discarding such suggestions, researchers propose using them in uncontrolled fields (e.g. MARC 653) to improve discovery in the library catalog.[19] This approach can enrich records with extra keywords for users to search, without breaking the rules of controlled vocabularies. Another way AI enhances recommendations is by leveraging semantic networks of terms. The LC "Exploring Computational Description" project built a prototype where an ML model identified an ebook's main topics and then"translated those topics into LCSH… terms," presenting catalogers with candidate headings along with their broader, narrower, and related terms from the LCSH hierarchy. This gave catalogers a menu of contextually related terms to choose from, likely improving the chance of picking the most precise headings. Such AI-powered suggestion interfaces not only speed up selection but also serve as a quality check – if a suggested term is too broad or too narrow, the interface can indicate that (as the LC prototype did with flags like

---

[17] Brador.
[18] Dobreski and Hastings, "AI Chatbots and Subject Cataloging."
[19] Chow, Kao, and Li, "An Experiment with the Use of ChatGPT for LCSH Subject Assignment on Electronic Theses and Dissertations."

"Too Broad" or "Wrong" next to suggestions).[20] In essence, AI acts as a preliminary cataloger, recommending subjects and even pointing to related vocabulary entries, which the human can then confirm or adjust. This collaboration can increase consistency across catalogers: given the same resource, the AI will suggest the same terms every time, reducing the variability that might occur between different people. Recommendation support from AI is especially valuable in specialized or interdisciplinary topics where catalogers might be less familiar – the AI can quickly draw connections to established headings in those areas. That said, studies caution that AI recommendations are "supportive" rather than definitive.[21] Librarians still need to apply their knowledge of cataloging rules (e.g. LCSH subdivision practice, priority of concepts) to finalize which terms to use or omit. When used thoughtfully, AI-driven recommendation systems in library science can enhance subject metadata by providing a strong initial suggestion set, improving the exhaustivity (more aspects covered) and consistency of subject terms across collections.[22] Ultimately, this leads to richer catalogs and potentially better resource discovery for patrons, as more relevant subjects are captured for each item.

**Integrating AI and LLMs with Controlled Vocabularies (LCSH)**

A key challenge and research focus in recent years is how to effectively integrate AI – particularly LLMs – with established controlled vocabularies like LCSH. Controlled vocabularies ensure consistency and authority in subject terms, so any AI-generated terms must align with these vocabularies to be truly useful in cataloging. Researchers have explored several approaches to achieve this alignment:

- **Prompt Engineering with Vocabulary Guidance:** One straightforward method is to craft prompts that guide an LLM toward using authorized terms. For example, Zhang, Wu, and Zhang had GPT-3.5 annotate research dataset records with subjects by providing the model with a predetermined list of subject categories (from the Australian New Zealand Research Classification scheme) in the prompt (Zhang, Wu, and Zhang (2023)).[23] By framing the task as

---

[20] Brador, "Could Artificial Intelligence Help Catalog Thousands of Digital Library Books?"

[21] Chow, Kao, and Li, "An Experiment with the Use of ChatGPT for LCSH Subject Assignment on Electronic Theses and Dissertations"; Brador, "Could Artificial Intelligence Help Catalog Thousands of Digital Library Books?"

[22] Koraljka Golub, "Automated Subject Indexing: An Overview," *Cataloging & Classification Quarterly* 59, no. 8 (November 29, 2021): 702–19, https://doi.org/10.1080/01639374.2021.2012311.

[23] Shiwei Zhang, Mingfang Wu, and Xiuzhen Zhang, "Utilising a Large Language Model to Annotate Subject Metadata: A Case Study in an Australian National Research Data Catalogue" (arXiv, October 17, 2023), https://doi.org/10.48550/arXiv.2310.11318.

*"choose the best fitting category from this list for the dataset"*, the LLM's output was effectively constrained to that controlled list. This few-shot prompting approach (sometimes including demonstration examples) showed "promising performance in automatic metadata annotation", though it struggled when records spanned multiple disciplines or when the subject needed more nuance than the coarse categories allowed.[24] The general insight is that carefully engineered prompts – for instance, explicitly instructing "provide 2–3 LCSH terms for this summary" or listing candidate terms to choose from – can leverage LLMs' knowledge while keeping them within bounds of a vocabulary. Prompt strategies are relatively easy to implement and require no model retraining, but they may hit limits in complexity (as seen by lower performance on very specific classification rules). Also, responses may be slightly different from time to time as they are next-token predictions which are subject to randomness. These minor differences can accumulate and lead to suboptimal results.

- **Fine-Tuning and Supervised Models:** Another approach is training or fine-tuning AI models on large sets of cataloged records so they learn to predict authorized terms directly. Prior to the rise of LLMs, library scientists experimented with machine learning classifiers for subject assignment – for instance, using algorithms like support vector machines, naïve Bayes, or neural networks trained on bibliographic records (Suominen, Lehtinen, and Inkinen (2022)).[25] Today, with transformers and LLMs, fine-tuning is being revisited. In a recent LC Labs experiment, catalogers assembled a training set of ~23,000 e-books with high-quality MARC records (including LCSH fields) to train AI models. Transformer-based models (which likely included fine-tuning on those records) achieved high accuracy on fields like title and author extraction, but for subjects the results were modest – an F1 accuracy of only ~35% for automatically assigning LCSH (Brador (2024)).[26] The low score is partly due to the "extreme multi-label" nature of subject indexing (each book can have multiple subjects, and there are hundreds of thousands of possible LCSH terms). Researchers learned that more training data is crucial: when they later expanded the training set to nearly 100,000 records, the model's accuracy improved, illustrating that fine-tuning needs very large, representative samples to capture a vocabulary as extensive as LCSH.[27] Fine-tuned models can eventually encode the controlled vocabulary into their parameters, potentially giving faster and more precise predictions than prompting alone. However, training such models is resource-intensive, and they require re-training or updates as vocabularies grow.

---

[24] Zhang, Wu, and Zhang.
[25] Osma Suominen, Mona Lehtinen, and Juho Inkinen, "Annif and Finto AI : Developing and Implementing Automated Subject Indexing," *JLIS*, no. 1 (2022), https://doi.org/10.4403/jlis.it-12740.
[26] Brador, "Could Artificial Intelligence Help Catalog Thousands of Digital Library Books?"
[27] Brador.

The LC experiment underscored that no single model hit the desired 95% accuracy threshold, reinforcing thathuman review remains necessary even with fine-tuning.[28] Nonetheless, as data availability and techniques improve, fine-tuning is expected to yield steady gains in aligning AI output with controlled terms.

- **Retrieval-Augmented Generation (RAG):** RAG is a technique where the model is supplemented with information retrieval – in this context, querying a knowledge base of authorized terms. Instead of relying solely on the LLM's internal knowledge, the system can search an authority file or vocabulary database (like LCSH in PDF files or index) for candidates, and then have the LLM incorporate or choose from those results. This was hinted at in some experiments: LC researchers mentioned using "vector 'searching' to create MARC fields and subfields" alongside LLMs (Brador (2024)). In practice, this could mean embedding an input text and retrieving the nearest neighbor terms from an LCSH embeddings index, then presenting those terms to the LLM to verify or format them properly. This kind of integration ensures the output terms exist in the vocabulary, dramatically improving precision. Retrieval-assisted approaches can thus combine an LLM's language understanding (finding relevant concepts in text) with the reliability of a controlled term database. Early indications are that this yields more acceptable headings for catalogers to choose from – effectively filtering out the hallucinations or non-standard terms that an LLM might invent. The trade-off is added complexity: setting up search indexes or APIs and possibly dealing with multiple results for a single concept (which the AI or cataloger must resolve). Still, many view RAG as a promising path to harness LLMs for subject indexing while grounding their output in authoritative data. Recently, Chow developed a RAG-based experiment for recommending Faceted Application of Subject Terminology (FAST).[29]

- **Function Calling or Using Tools for Validation:** A very recent innovation in LLM technology is *function calling*, where the model can be constrained to return data in a structured format and trigger predefined functions. In a cataloging context, one could define a function (e.g. get_LCSH(term)) that looks up a term in the LCSH authority file. The LLM, when asked to assign subjects, would be required to output a call to get_LCSH() with potential subject keywords, which the system then resolves to authorized headings. This ensures the final output consists of valid controlled terms. While we haven't yet seen a peer-reviewed library science study explicitly using function calling for subject terms, the technique is conceptually aligned with the needs of cataloging. By having the AI output *structured JSON or MARC snippets* with identifiers for subject headings, for instance, it becomes easier to automatically validate and integrate those suggestions into catalogs. Chow, Kao, and Li noted that incorporating tools to check term validity can mitigate some issues of LLM-generated subjects – function calling is precisely the kind of mechanism to implement such checks in real-time.[30] We can expect that future cataloging applications will utilize this feature of LLM APIs to enforce vocabulary control, effectively bridging free-text generation and strict cataloging rules.

---

[28] Brador.

[29] Eric H. C. Chow, "Exploring the Future of Library Cataloging with AI and Multilingual Embeddings," *The Digital Orientalist* (blog), May 9, 2025, https://digitalorientalist.com/2025/05/09/exploring-the-future-of-library-cataloging-with-ai-and-multilingual-embeddings/.

**Solution: Validating LLM-Generated LCSH Terms with the LOC ID Service**

To address the challenges of accuracy and consistency in AI-generated subject headings, we have developed a three-stage solution centered on validating these suggestions against the Library of Congress (LOC) Linked Data Service. This approach aims to make use of the analytical power of LLMs while grounding their outputs in the authoritative vocabulary of the LCSH.

*The Core Idea: An Iterative Refinement Process*

The fundamental concept of our solution is an iterative process where the LLM's initial suggestions are cross-referenced with the LOC ID Service, and the feedback from this validation is used by the LLM to produce a more accurate and reliable set of final LCSH terms. (Figure 1)

*First Stage: LLM Suggesting LCSH Terms*

1. **Bibliographic Input**: The user provides bibliographic information about a work (e.g., title, author, abstract, table of contents, and potentially images of the cover or title page) to an LLM-powered chatbot or interface.
2. **Initial LLM Suggestion**: The LLM processes this input and generates an initial list of candidate LCSH terms that it deems relevant to the work.

*Second Stage: Validating Suggested LCSH Terms*

---

[30] Chow, Kao, and Li, "An Experiment with the Use of ChatGPT for LCSH Subject Assignment on Electronic Theses and Dissertations."

3. **LOC ID Service Validation**: These candidate terms are then programmatically queried against the Library of Congress Linked Data Service (specifically, services like [id.loc.gov/authorities/subjects/suggest2](id.loc.gov/authorities/subjects/suggest2) or similar APIs). This step verifies if the terms are valid LCSH entries and can retrieve additional information such as authorized forms, URIs, and related terms.
4. **Feedback to LLM**: The results from the LOC ID Service (e.g., validation status, official forms of headings, alternative suggestions, related terms) are returned and provided as new context to the LLM.

*Third Stage: Finalizing Suggestions*

5. **Finalized LLM Recommendations**: Enriched with this validation feedback, the LLM refines its initial suggestions. It can confirm valid terms, correct misformatted ones, replace non-standard terms with authorized equivalents, and potentially leverage related terms to improve the comprehensiveness of its final recommendations.
6. **Output to User**: The finalized, validated list of LCSH terms is presented to the user, often accompanied by justifications and direct links to the corresponding authoritative entries in the LOC ID Service. By following these links, users can readily explore existing works associated with each specific term.

This iterative loop ensures that the LLM's suggestions are not just plausible but are also aligned with the established controlled vocabulary, thereby enhancing the quality of subject cataloging.

*Deployment Approaches*

We have explored and implemented this validation solution through three distinct technical approaches, each tailored to different LLM interaction paradigms:

1. **Middleware API Service for ChatGPT Custom GPTs (Function Calling)**:[31]
   - **Concept**: Function calling, a protocol developed by OpenAI, enables its LLMs to utilize external tools and services during the inference process. This functionality is available for both the OpenAI API and custom GPTs. Since custom GPTs exclusively support the OpenAPI schema, a backend API service was created to act as an intermediary between the custom GPTs and the LOC Linked Data Service. This middleware API not only facilitates communication but also calculates and provides similarity scores between the subject terms suggested by the LLM and the results retrieved from the LOC Linked Data Service.[32]

---

[31] "LCSH API - Swagger UI," accessed June 17, 2025, https://lcsh.098484.xyz/docs.

- **Mechanism**: A custom GPT named "LCSH Recommendation for East Asian Librarians" was developed.[33] This GPT is designed to invoke a specific function whenever it needs to validate subject terms. When validation is required, the GPT triggers a function call that is directed to our middleware API. This API then takes the suggested subject term(s) and queries the Library of Congress ID Service. The API processes the results from the LOC ID Service and sends a structured response back to the custom GPT. The GPT uses this structured data to refine its final recommendations. This setup allows ChatGPT to leverage external, real-time information from the LOC ID Service, enhancing its capabilities beyond its inherent training data.
- **Pros and Cons**: Leverages external, up-to-date LOC ID Service information, significantly expanding ChatGPT's subject term suggestion capabilities beyond its internal training data. Restricted to paid ChatGPT subscribers; geographic limitations may hinder use in certain regions.

2. **Google Chrome Extension with Gemini API Integration**:[34]
    - **Concept**: A browser extension provides a user interface for inputting bibliographic data (including image uploads for multimodal analysis) and directly leverages the Gemini API for LCSH suggestions.
    - **Mechanism**: The extension's JavaScript code orchestrates the process. After the user inputs data, the extension calls the Gemini API (e.g., Gemini 2.0 Flash) to get initial LCSH suggestions. The extension then takes these suggestions and makes client-side requests to the LOC ID Service. The validation results from LOC are then compiled and can be sent back to the Gemini API in a subsequent prompt, asking it to refine the suggestions based on this new, verified information. The extension manages the display

---

[32] For full codebase of the API, "Kltng/Lcsh-Validation-Api: A FastAPI-Based API Service That Provides Library of Congress Subject Headings (LCSH) Recommendations Based on Input Keywords. The Service Scrapes the Library of Congress Website, Performs Similarity Comparisons, and Returns the Most Relevant LCSH Terms.," GitHub, accessed June 17, 2025, https://github.com/kltng/lcsh-validation-api.

[33] https://chatgpt.com/g/g-67c3bb30ceb881919481c8942ce75766-lcsh-recommendation-for-east-asian-librarians

[34] https://chromewebstore.google.com/detail/lcsh-recommendation-tool/gdabagnphjdcojgnogaiapkiheipobnp

of initial suggestions, LOC results, similarity scores, and final recommendations. We provide a YouTube video for introducing the Chrome Extension.[35]
- **Pros and Cons:** This browser-integrated tool offers convenience for catalogers working with online resources and enhances context through multimodal input (text and images). It does not require a paid subscription as Google Gemini provides a free tier with enough of quota. However, the Chrome extension may require periodic updates to address security vulnerabilities and implement functional improvements.

3. **Model Context Protocol (MCP) Server for Validation**:
    - **Concept**: Model Context Protocol (MCP) is a standard developed by Anthropic for exchange data with external resources and services. An MCP server is implemented to expose the LCSH validation functionality as a standardized "tool" or "resource" that MCP-compatible LLMs (such as Claude) can utilize.
    - **Mechanism**: The MCP server (lcsh-mcp-server built with Python) defines a tool (e.g., search_lcsh) that accepts a query term. When an LLM connected to this MCP server needs to validate an LCSH term, it invokes this tool. The MCP server handles the communication with the LOC Linked Data Service (e.g., the suggest2 API) and returns the results to the LLM in a format it can understand. The LLM then uses this information to confirm or refine its subject heading suggestions. We developed a MCP server called Cataloger-MCP.[36]
    - **Pros and Cons**: MCP enhances interoperability by enabling various LLMs to consistently access specialized tools and data sources, streamlining the integration of external knowledge into their reasoning. This makes it model-independent. Currently, setting up an MCP server with MCP hosts can be difficult for novice users, but this process will likely become simpler as better tools are developed.

Each of these deployment methods successfully implements the core validation loop, demonstrating the versatility of the proposed solution across different AI platforms and user environments.

**Conclusion**

In summary, this article presents a new solution to address the challenges of subject cataloging in library work. These long-existing problems, such as the complexity of the LCSH system, the time-consuming nature of using subdivisions, and the resulting substantial cataloging backlog, continue to put pressure on library workflows. This solution, based on LLM-generated LCSH terms and automated validation through LOC Linked Data Service,  holds promise to significantly expedite subject cataloging and help alleviate the current situation.

---

[35] *LCSH Recommendation Tool: Introduction*, 2025, https://www.youtube.com/watch?v=aQ9vxCNn4jQ.
[36] Kwok-leong Tang, "Kltng/Cataloger-Mcp," Python, May 30, 2025, https://github.com/kltng/cataloger-mcp.

Recently, we have shared two of the tools introduced above, the Custom GPT "LCSH Recommendation Tool for East Asian Librarians" and the browser extension "LCSH Recommendation Tool", with a broad community of cataloging librarians and encouraged them to experiment with them. The feedback received was enthusiastic.  These tools are found to be effective in many ways for day-to-day cataloging, including improving work efficiency, reducing repetitive manual operations, providing real-time validation and learning support, etc. One notable advantage is that they enable catalogers to focus more on content analysis instead of being constrained by tedious rule proofreading. Another valuable attribute lies in the automated quality control step that can help identify invalid or inconsistent subject term strings. Overall, these tools show great potential for large-scale metadata creation and optimization projects, thereby supporting the goals of reducing massive backlogs and processing complex collections.

Last but not least, it is important to emphasize that these AI tools are not designed to replace human catalogers. Rather, they are built to work alongside us like an assistant. Catalogers should always play a pivotal role in the whole process, using their professional judgment, enabling consistency, and reinforcing high-quality subject access. While automation helps accelerate workflow processes, it is the expertise of catalogers that ultimately ensures meaning, nuance, and accuracy in bibliographic work. Looking ahead, the continued development of these tools will rely on the supervision of catalogers and need to adapt to the evolving cataloging standards. A dynamic and deepening human-AI collaboration with active and broad participation of catalogers will be the key to shaping future cataloging practices with efficiency and integrity.

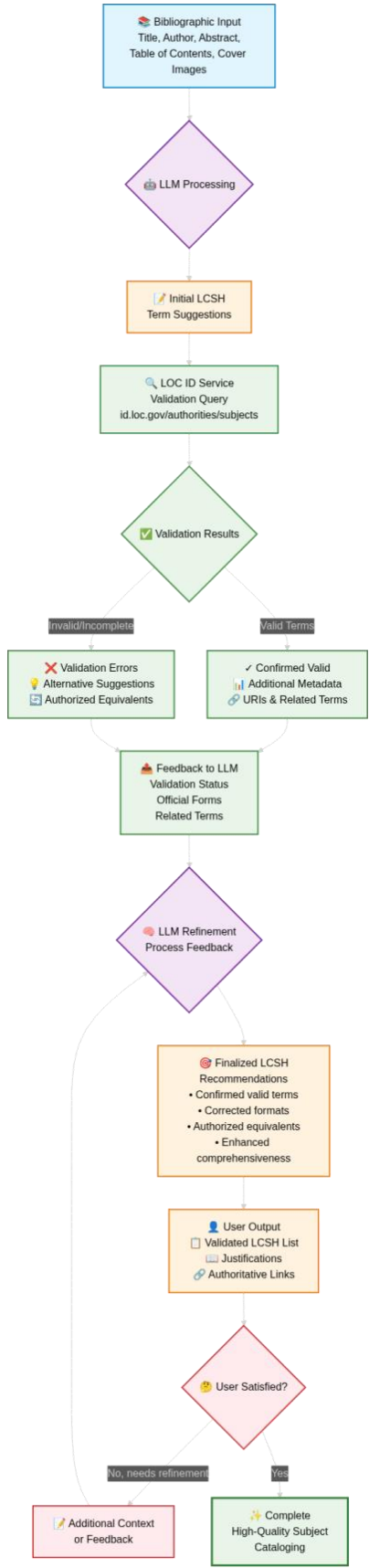

**Figure 1. The process of validation.**

**AI Statement**

The authors used generative AI in the following ways in this article:

**Code assistance and debugging**: The authors used GitHub Copilot, Cursor, and Windsurf for code debugging in building the LCSH API, the Chrome extension, and the MCP server.

**Language proofreading:** The authors used Google Gemini in Google Doc and Microsoft Copilot in Microsoft Word to enhance language accuracy and clarity during proofreading..